# Fission Enhanced diffusion of uranium in zirconia

N. Bérerd, A. Chevarier, N. Moncoffre,

Institut de Physique Nucléaire de Lyon , 4, rue Enrico Fermi , 69622 Villeurbanne Cedex, France,

Ph. Sainsot,

Institut National des Sciences Appliquées, UMR 5514, 69621 Villeurbanne Cedex, France

H. Faust,

Institut Laue Langevin, BP 156, 38042 Grenoble Cedex 9, France,

H. Catalette,

EDF R&D, Site des Renardières, 77818 Moret sur Loingt, Cedex, France.

**Abstract**

This paper deals with the comparison between thermal and Fission Enhanced Diffusion (FED) of uranium into zirconia, representative of the inner face of cladding tubes. The experiments under irradiation are performed at the Institut Laue Langevin (ILL) in Grenoble using the Lohengrin spectrometer. A thin $^{235}UO_2$ layer in direct contact with an oxidized zirconium foil is irradiated in the ILL high flux reactor. The fission product flux is about $10^{11}$ ions cm$^{-2}$ s$^{-1}$ and the target temperature is measured by an IR pyrometer. A model is proposed to deduce an apparent uranium diffusion coefficient in zirconia from the energy distribution broadening of two selected fission products. It is found to be equal to $10^{-15}$ cm$^2$ s$^{-1}$ at 480°C and compared to uranium thermal diffusion data in $ZrO_2$ in the same pressure and temperature conditions. The FED results are analysed in comparison with literature data.

## 1. Introduction

During reactor operation, the fuel cladding inner surface is oxidised. This layer is non homogeneous first in thickness, second in composition (presence of hydrides, amorphous precipitates) and finally its structure is strongly strained [1]. In addition, it is contaminated with fission products and actinides [2]. Actinide contamination is due to successive recoil effects induced by fission products and alpha emission. Since the Pressurised Water Reactor (PWR) most common fuel is $UO_2$, the probability of contamination by uranium is largely predominant compared to other actinides and, in addition, uranium converts to other actinides, mainly Pu and Am by successive (n,γ) reactions and β emission. One important question to answer concerns the contribution of uranium diffusion under irradiation in the cladding contamination. The understanding of the phenomena occurring during reactor operation and their consequences are crucial to model the safety of dry disposals.

The aim of this paper is to investigate the migration behaviour of uranium in $ZrO_2$ under fission product irradiation. The present experiment was performed using the high neutron flux of the Institut Laue Langevin (ILL) in Grenoble. The diffusion coefficient under irradiation is compared to thermal data obtained in the same pressure and temperature conditions in order to evaluate the damage role.

## 2. Fission Enhanced diffusion

A major drawback of this experiment is that it is impossible to collect the sample after irradiation, because of the very high radioactivity level. However, it has been shown [3] that around 500°C, the stable zirconia structure is monoclinic and that under irradiation, a given proportion of tetragonal zirconia is observed. A recent work using 50 MeV Xe irradiation [4] shows that under at 480°C and $5 \times 10^{-3}$ Pa (which are the same temperature and pressure conditions as those used in this study), zirconium oxidation is accelerated and the obtained zirconia structure is a mixture of tetragonal phase (50%) and monoclinic phase (50%). This composition is stable for xenon fluences close to the FP fluence used in this experiment.

*2.1 Experimental set up*

The ILL is equipped with a high neutron flux nuclear reactor ($\Phi = 5 \times 10^{14}$ n cm$^{-2}$ s$^{-1}$) and we have used the H9 beam line, on which the Lohengrin mass spectrometer is located. The residual pressure in the beam tube is close to $5 \times 10^{-3}$ Pa. The line is composed of three main parts: the target, the Lohengrin spectrometer and an ionisation chamber as detector. The target is made of a titanium support covered with a thin platinum deposit on which a 300 μg cm$^{-2}$ uranium oxide layer enriched up to 98% with $^{235}$U has been deposited. A 3.1 μm thick x 3.5 cm$^2$ zirconia foil (obtained from the irradiation enhanced oxidation of a 2 μm thick zirconium foil) is in contact with the $^{235}UO_2$. Figure 1 presents a scheme of the target irradiation set up. This device is positioned in the ILL thermal neutron flux. The corresponding fission fragment (FF) flux generated by the $^{235}UO_2$ film is estimated to be $10^{11}$ particles cm$^{-2}$ s$^{-1}$. Uranium diffusion in the zirconia target will occur under neutron and FF irradiation. These irradiation conditions are very close to Matzke's experiments [5] which demonstrate that the fission products are the most efficient in the diffusion enhancement. In the following, we will use the term Fission Enhanced Diffusion (FED).

The principle of our experiment is the measurement of the energy distribution evolution of a selected FF using the Lohengrin spectrometer. High energy FF are emitted from the $^{235}$U thermal fission and pass through zirconia before being detected. Their mass and kinetic energy are analysed in three successive magnetic and electric fields. At the exit point, all FF having the same M/q ratio (M is the FF mass and q is its ionic charge state) and the same velocity are selected. The separated fragments are identified using a high resolution ionisation chamber [6]. We have chosen to detect the masses 90 and 136 with the most probable charge state (q=18). At regular time intervals, the energy spectra of the chosen fission fragments are measured. The spectrum obtained at t=0, representing the relative FF intensity versus its kinetic energy, is shown in figure 2. This spectrum is the result of an energy scan. The number of events, n, corresponds to the energy, E, and to an acquisition time, t. It is given by N :

$$N = \frac{n}{E\,t} \qquad (1)$$

It allows to express the number of events as function of time and to take into account the fact that the energetic acceptance of the spectrometer is proportional to $\frac{1}{E}$. In addition, all the spectra are normalised to a relative intensity equal to 1 at the spectra maximum.

The target temperature was measured during irradiation using a specific infra-red pyrometer, which allowed the determination of the target temperature at a distance of 15 m, on a 8x4 mm$^2$ spot and with an accuracy of 10°C. Such an experiment has been previously described [7, 8]. The temperature measurement required the use of two gold mirrors in order to avoid the direct view of the pyrometer camera towards the high neutron flux. Figure 3 displays the general set up. In our experiment, temperature was measured twice a day; it remained constant at a mean value of 480°C.

*2.2 Results*

Figure 4 displays the evolution of the FF kinetic energy distribution for three diffusion times (t=0, t=380 h, t=546 h) and at 480°C. Each distribution is fitted with gaussian-like curves. We observe that the kinetic energy distributions broaden while the mean kinetic energy value remains constant (50 MeV for M=90 and 20 MeV for M=136). Only the right part of the distribution, which is correlated to uranium diffusion in zirconia, is considered since the low energy part corresponds to diffusion in the platinum deposit of the target support.

*2.3 Modelling of the diffusion process under irradiation*

In order to deduce an apparent diffusion coefficient of uranium in zirconia, we have simulated the evolution of the energy spectra as a function of time. In this model, uranium diffusion follows the Fick's second law according to:

$$\frac{\partial C}{\partial t} = - D^* \frac{\partial^2 C}{\partial x^2} \qquad (2)$$

where C is the atomic concentration of the diffusing uranium, $D^*$ is the diffusion coefficient under irradiation ($cm^2 \, s^{-1}$) and x the diffusion depth (cm). The solutions depend on boundary conditions which are the following:

for  x > 0, C = C(x,t),

  x = ∝, C(∝,t) = 0,

  t = 0, C(x,0) = 0.

  x ≤ 0, C = $C_S$ = 1/3. The number of diffusing uranium atoms is negligible compared to the total amount of uranium in the deposit. So, in first approximation, $C_S$ is constant and equal to the atomic uranium concentration in $UO_2$.

These boundary conditions imply a uranium diffusion study within concentration gradients. They lead to the following analytical solution :

$$C(x, t) = C_S \, \text{erfc} \left( \frac{x}{2\sqrt{D^* \, t}} \right) \qquad (3)$$

where t is the irradiation time (s).

Figure 5 gives a schematic representation of our approach which is based on the following conditions : before diffusion, the thin $UO_2$ layer (constant uranium concentration) corresponds to a gaussian energy distribution of FF (fig. 5a). As uranium diffuses, the FF distribution broadens (fig. 5b). For a given irradiation time t, a $D^*$ value is supposed from which, according to equation (3), one can calculate an uranium diffusion profile. This calculated profile is discretised in elementary depths dx in which the uranium concentration is supposed to be constant. Each elementary part is associated to a Gaussian-like distribution with the same half-width but with a maximum energy deduced from the energy loss calculations (fig. 5c). At a given time t, the FF energy profile is rebuilt by summing the effects of all elementary contributions. The correct value of $D^*$ is obtained by fitting the result of the model to the experimental spectrum. An illustration of such a simulation is presented in figure 6 for the mass 90. We observe that uranium diffuses up to 1.6 µm in depth after an irradiation time of 546 hours. We have estimated with the SRIM code [9] the amount of uranium atoms recoiling by FF elastic diffusion. The calculation result shows that beyond 0.2 µm, the recoiling uranium yield is negligible.

Finally, the apparent uranium diffusion coefficient in $ZrO_2$ under irradiation deduced from this model at 480°C and for a flux of $10^{11}$ FF cm$^{-2}$ s$^{-1}$ has been determined to be equal to $(1.0 \pm 0.1) \times 10^{-15}$ cm$^2$ s$^{-1}$. The error bar is determined from the diffusion model fit.

### 3. Thermal diffusion

To compare the results under irradiation with thermal data, a diffusion study was performed by coupling ion implantation and Rutherford Backscattering Spectrometry. Zirconium oxide was obtained by annealing polycrystalline zirconium foils in air at 450°C during 5 hours. In these conditions, the zirconia layer in the Zr substrate reaches 1.5 µm in thickness. These samples were then implanted at the Nuclear Physics Institute of Lyon with $^{238}U^{2+}$ ions of 800 keV energy and with a fluence of $10^{16}$ ions cm$^{-2}$ s$^{-1}$. The uranium range calculated with SRIM is 100 nm and the maximum U concentration is about 1 at.%, in agreement with RBS analysis. Complementary grazing angle X-ray diffraction analysis shows an increase of the tetragonal phase up to 36 % induced by implantation [8] in agreement with literature data [10, 11]. Annealings were performed at 800°C under primary vacuum ($7.5 \times 10^{-1}$ Pa). The study could not be performed in a secondary vacuum because of the zirconia dissolution in such conditions [12].

RBS spectra registered with 3 MeV alpha particles do not show any broadening of the U distributions after a 23 h annealing at 800°C as shown in figure 7. This means that in our experimental conditions, no uranium diffusion can be detected. Therefore, the thermal diffusion coefficient $D_{th}$ is lower than $10^{-18}$ cm$^2$ s$^{-1}$.

## 4. Discussion

We have shown that thermal diffusion of uranium in $ZrO_2$ is extremely low (<$10^{-18}$ $cm^2$ $s^{-1}$ at 800°C). These results are in agreement with data concerning cation transport in zirconia, and which are collected in the review paper of Kilo et al. [13]. Under FF irradiation, the diffusion coefficient strongly increases; it reaches $10^{-15}$ $cm^2$ $s^{-1}$ under a FF flux F of $10^{11}$ particles $cm^{-2}$ $s^{-1}$ at a temperature of 480°C. Therefore irradiation defects are predominant on the diffusion mechanism.

As known from the theory of the diffusion under irradiation [14, 15], irradiation increases the point defect concentration in the material, and this increase influences diffusion in the solid. $D^*$ can be written as :

$$D^* = D_{irr} + D_{th} \qquad (4)$$

with $D_{irr}$ and $D_{th}$ respectively the diffusion under irradiation and the thermal diffusion contributions [14]. From results obtained in parts 2.3 and 3, it appears that $D_{th}$ is negligible in regards to $D_{irr}$, and so $D^*$ is very close to $D_{irr}$. Therefore, we consider that the diffusion of uranium in $ZrO_2$ is not thermally activated and depends on the FF flux only [16]. We can estimate the uranium diffusion coefficient $D_{irr}$ in $ZrO_2$, generated by irradiation induced FF ballistic effects. As usually considered [17]:

$$D_{irr} = \frac{1}{6} \Gamma R^2 \qquad (5)$$

where R is the root-mean square displacement of a uranium atom in the collision cascade and $\Gamma$ is the jump rate proportional to the atomic displacement rate F in dpa $s^{-1}$ ($\Gamma = \alpha F$, where $\alpha$ corresponds to the number of atomic jumps per displacement).

In order to determine F, it is necessary to estimate the number of defects created at ILL. Only defects induced by FF will be taken into account in this evaluation. The mass and kinetic energy distributions of fission products are well known and it is usual to consider that the ion with mass 117 at 0.7 MeV/u is representative of fission products. Hence, in SRIM calculations, we have considered the $^{117}In$ isotope with a kinetic energy equal to 82.5 MeV. Moreover, the FF emission is isotropic and the range of the FF in the foil depends on the emission angle θ. We have calculated the number of dpa created as a function of depth for emission angles varying from 0° to 88°. An average distribution is deduced from those contributions. It corresponds to a mean F value equal to $6.4 \times 10^{-5}$ dpa $s^{-1}$.

In order to evaluate R, we have been done a SRIM full cascade calculation and we have obtained that the mean recoil energy value of uranium in $ZrO_2$ corresponds to R equal to 15 nm.

The $D^*$ experimental values were reproduced by using equation (5) in which $\alpha$ is equal to 40. This high value puts in evidence that atomic displacements only, cannot explain the radiation enhanced diffusion. It is comparable to the results of Müller [17] who introduced an $\alpha$ value of 125 and concluded to a significant influence of radiation induced sinks (dislocations, grain boundaries, vacancy and interstitial clusters,…) generated along the heavy ion trajectory.

We have compared our results with those of Matzke, who measured uranium and plutonium diffusion in $UO_2$ in reactor irradiations [5, 16, 18]. His experimental conditions are close to ours, since he follows the behaviour of thin $^{233}UO_2$ or $^{238}PuO_2$ tracer layers deposited on 1 mm thick $UO_2$ single crystals. To interpret his results, Matzke has considered that the whole damage, including displacement cascades as well as changes in the material physical properties, is produced along the fission product trajectory. He has shown that for temperatures lower than 1000°C, the $D^*$ coefficient of U and Pu in $UO_2$ is not thermally activated and is proportional to the fission rate. He normalised these data to a rate value of $5 \times 10^{12}$ fissions $cm^{-3}$ $s^{-1}$, and the corresponding $D^*$ value of U and Pu in $UO_2$ is around $7 \times 10^{-17} cm^2 s^{-1}$ [16].

In order to go further in the comparison, we have estimated the fission product flux $\Phi$ reaching the $^{233}UO_2$ surface layer in Matzke's experiment using the Hocking expression [19].

$$\Phi = 2 \, f \, d \qquad (6)$$

where, f is the fission rate, d (8 μm) is the mean fission product range in $UO_2$ and the factor of 2 arises from allowing for two fission fragments per fission. We found a value of $8 \times 10^9$ FF $cm^{-2}$ $s^{-1}$. In our experiment, $D^*$ of U in $ZrO_2$ equal to $10^{-15}$ $cm^2$ $s^{-1}$ is induced by a FF flux equal to $10^{11}$ FF $cm^{-2}$ $s^{-1}$. Assuming that the FED $D^*$ of U in $ZrO_2$ is proportional to the FF flux, then, our $D^*$ value normalised to Matzke's conditions is equal to $8 \times 10^{-17}$ $cm^2$ $s^{-1}$. These data are summarised in table 1. The similar behaviour of these 2 insulators $UO_2$ and $ZrO_2$ towards uranium FED can be explained by similar processes of defect evolution induced by bombardment with heavy ions [20].

## 5. Conclusion

We have shown that under FF irradiation, uranium diffused in zirconia, with a diffusion coefficient at 480°C equal to $10^{-15}$ cm$^2$ s$^{-1}$. To compare this result with uranium FED data in $UO_2$, we have considered that the damage on the whole FF range influences the surface diffusion. In such conditions, the diffusion coefficients are in good agreement, which put in evidence that under FF irradiation, U has the same behaviour in $UO_2$ and $ZrO_2$. This is corroborated by the large and similar value of the experimental D* and the theoretical $D_{irr}$ coefficients. This last coefficient is weakly dependent on atomic displacements in the collision cascade but is largely influenced by radiation induced sinks which enhance the uranium mobility.

**Acknowledgements**

The authors are very grateful to N. Chevarier, H. Jaffrézic and Y. Pipon for fructuous discussions. They also thank J.C. Duclot for his help during ILL experiments.

Figure Captions

Figure 1: Scheme of the target irradiation set up.

Figure 2: Example of a kinetic energy distribution of fission fragments with M = 90, registered at the beginning of uranium diffusion.

Figure 3: Schematic representation of the experimental set up on the PN1 beam line at ILL, including the temperature measurement.

Figure 4: Evolution of fission fragment kinetic energy distribution of mass M= 90 for three diffusion times (0, 380 h, 546 h) . Experimental data have been fitted by gaussian-like curves.

Figure 5: Schematic representation of the diffusion model showing the relation between the uranium diffusion and the energy distribution of a selected FF. (a) Initial conditions, (b) Intermediate time, (c) Scheme of the resolution method.

Figure 6: Result of the diffusion simulation: (a) uranium profile for $D=10^{-15}$ cm$^2$ s$^{-1}$ and (b) the fitted curve (dashed lines) of the final M = 90 energy distribution.

Figure 7: Uranium distribution profiles in zirconia as a function of annealing time at 800°C (t= 0, 8 h, 23 h).

Table Caption

Table 1: Diffusion coefficients under irradiation and dependence with the FF flux.

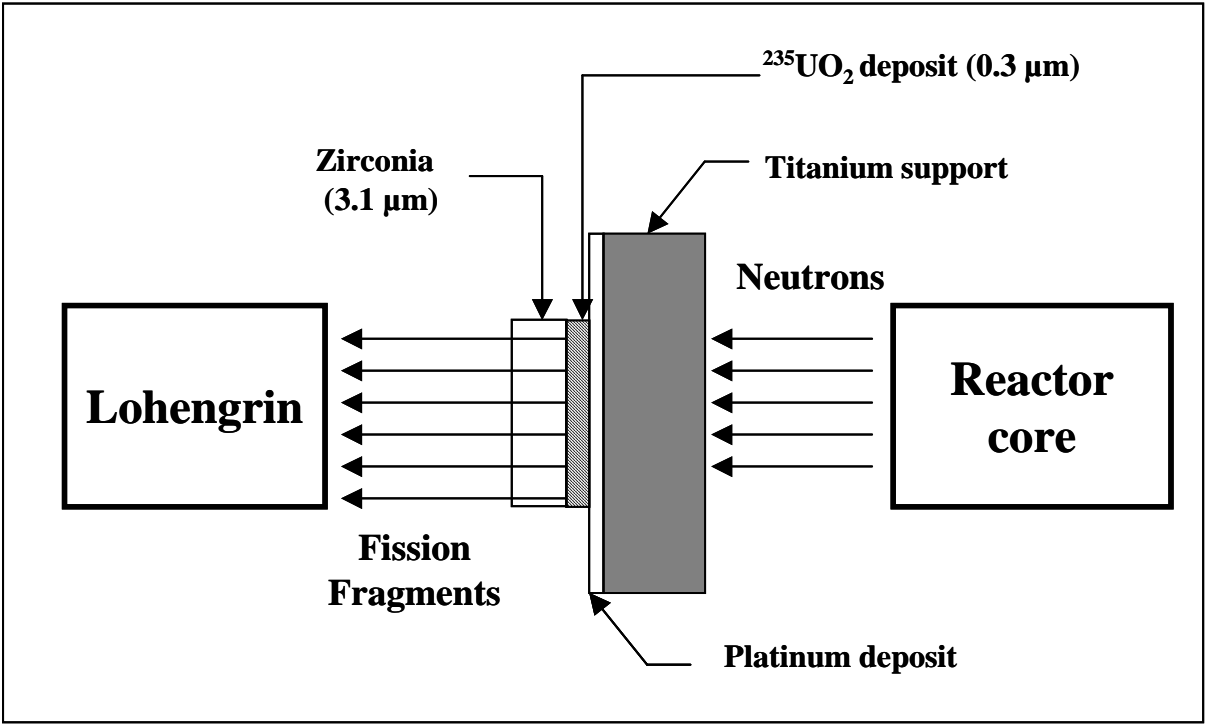

**Figure 1**

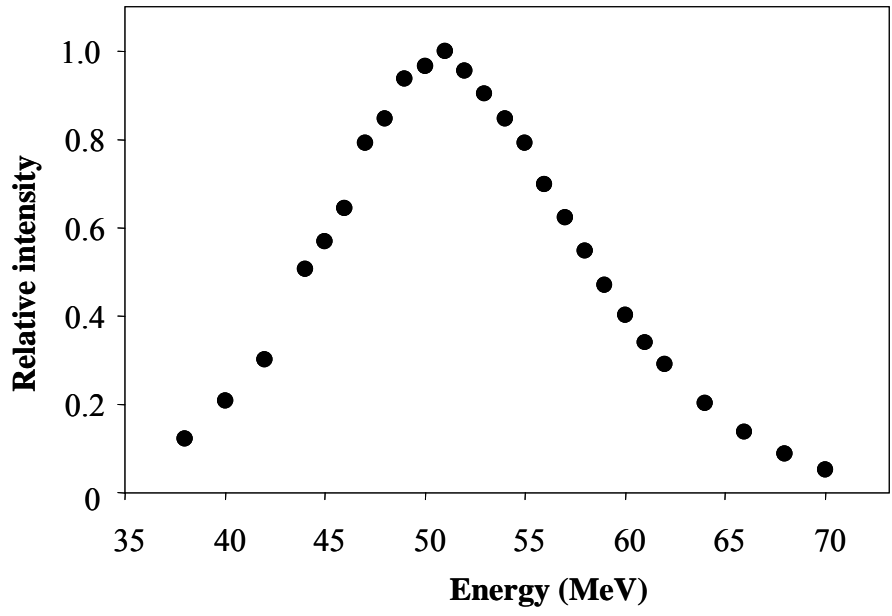

**Figure 2**

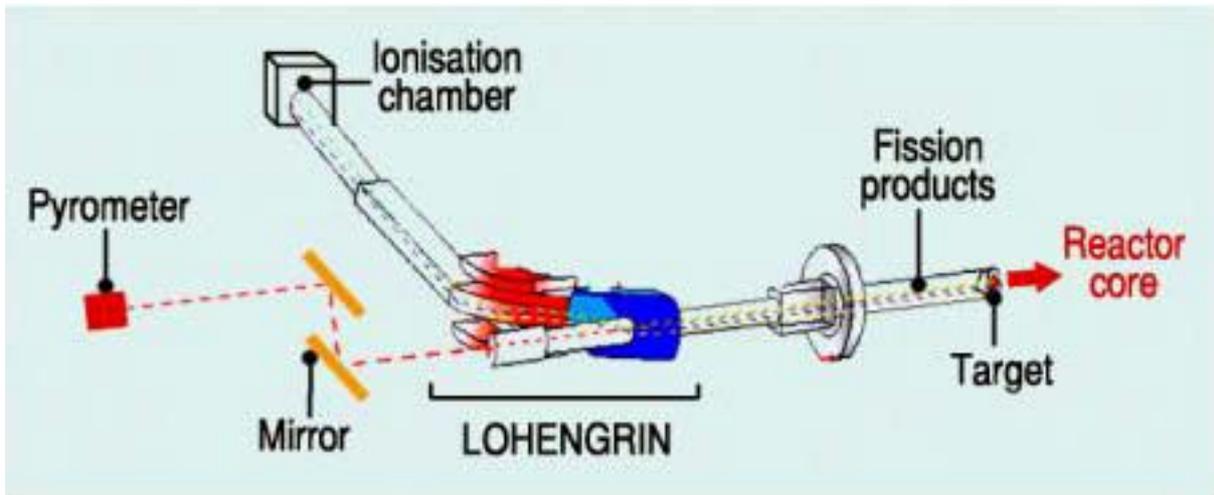

**Figure 3**

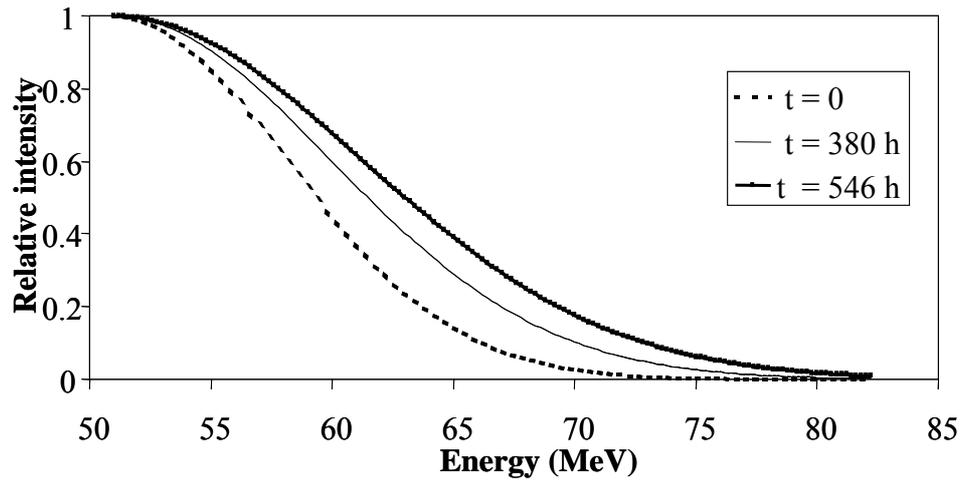

**Figure 4**

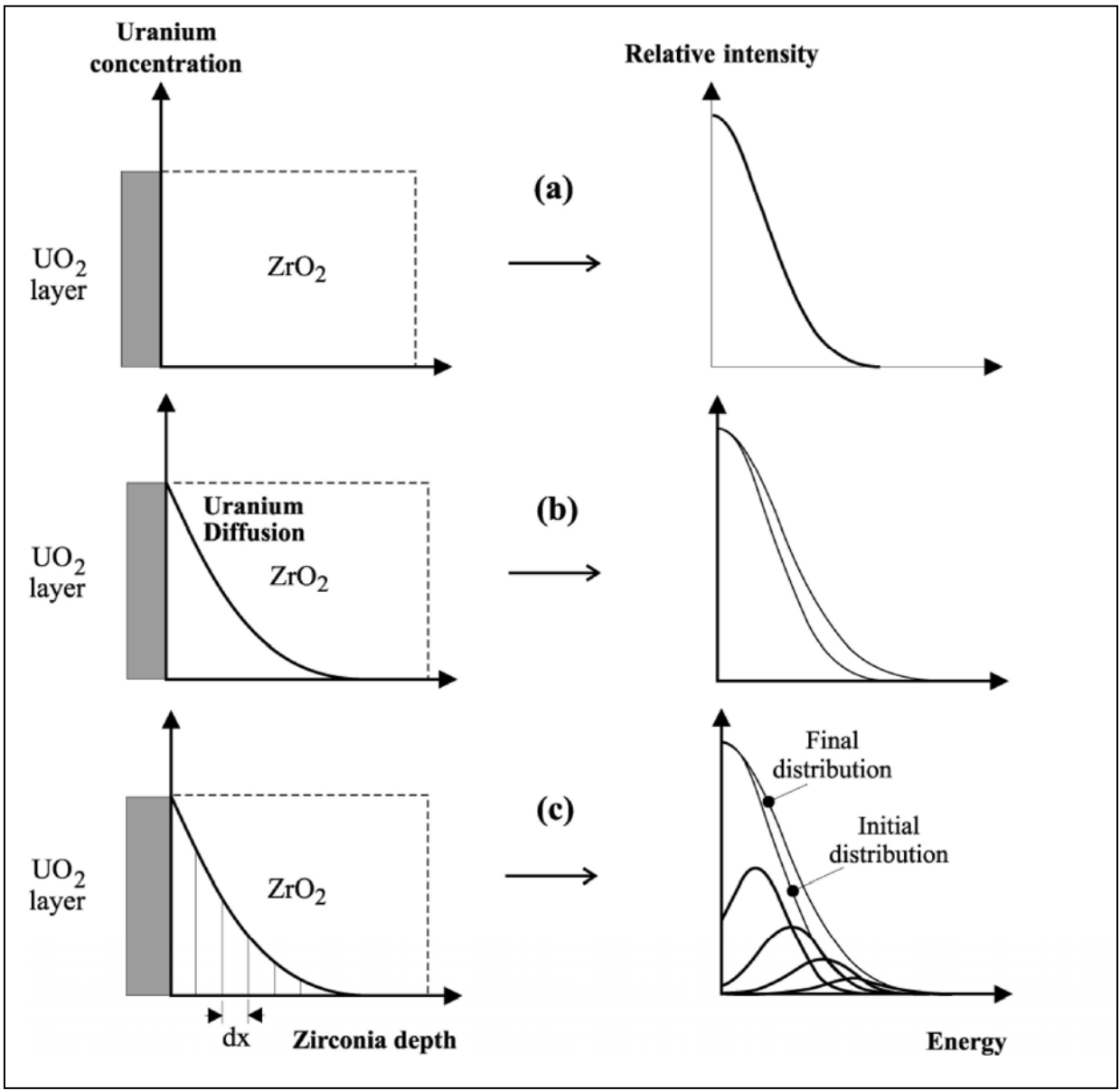

**Figure 5**

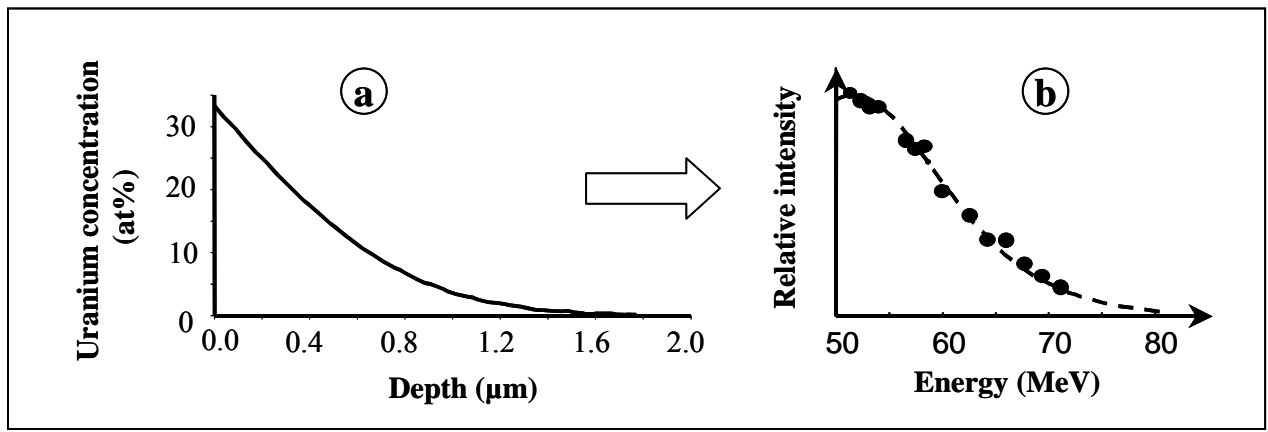

**Figure 6**

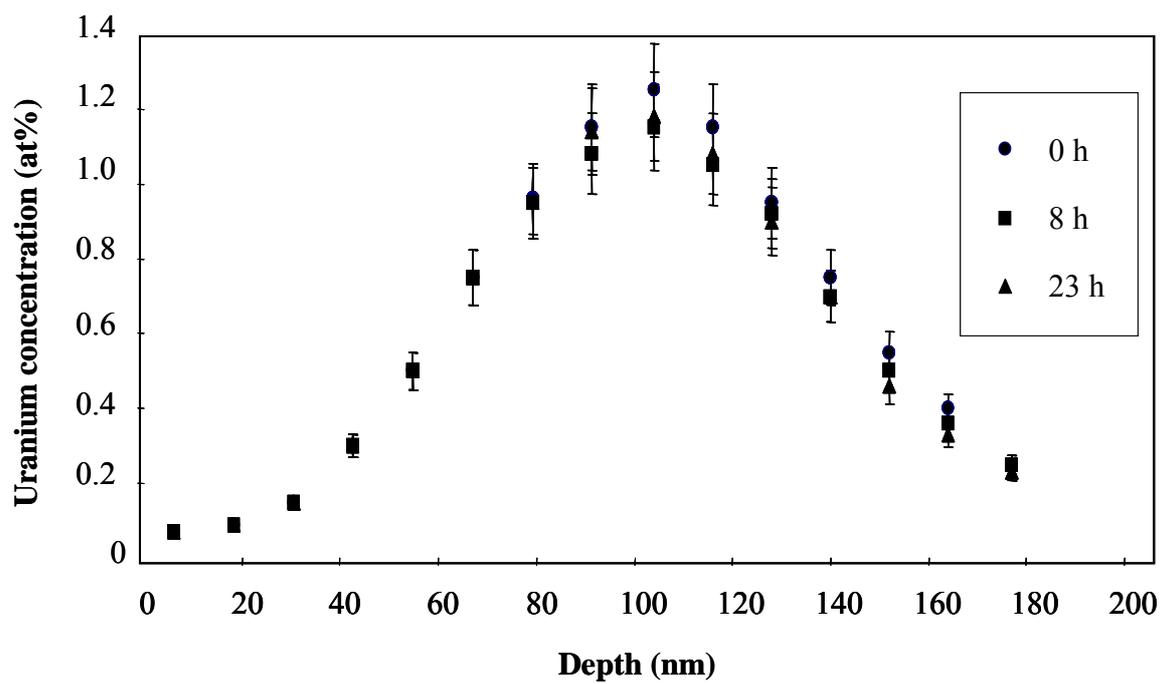

**Figure 7**

| Authors | F (FF cm$^{-2}$ s$^{-1}$) | D* (cm$^2$ s$^{-1}$) |
|---|---|---|
| D$^*$ (U in UO$_2$) Matzke [16] | 8x10$^9$ | 7x10$^{-17}$ |
| D$^*$ (U in ZrO$_2$) This experiment | 10$^{11}$ | 10$^{-15}$ |
| This experiment normalised to a 8x10$^9$ flux value | 8x10$^9$ | 8x10$^{-17}$ |

**Table 1**